\documentclass[preprint]{aastex631}
\usepackage{amsmath}
\usepackage{amssymb}
\usepackage{textcomp}

\submitjournal{AJ}

\shorttitle{Adaptive algorithm for detecting double stars}
\shortauthors{Sazhin et al.}
\graphicspath{{./}{figures/}}

\begin{document}

\title{An adaptive algorithm for detecting double stars in astrometric surveys}

\author{Mikhail V. Sazhin}
\affiliation{Sternberg Astronomical Institute of Lomonosov Moscow State University,\\
	Universitetsky pr., 13, Moscow 119234, Russia}

\author[0000-0002-3428-0106]{Valerian Sementsov}
\affiliation{Sternberg Astronomical Institute of Lomonosov Moscow State University,\\
	Universitetsky pr., 13, Moscow 119234, Russia}

\author{Sergey Sorokin}
\affiliation{Tver State University, Zhelyabova 33, Tver, Russia}

\author{Dan Lubarskiy}
\affiliation{Scienteco, Inc., Boston, MA}

\author{Alexander Raikov}
\affiliation{Institute of Control Sciences, Russian Academy of Sciences}



\begin{abstract}
The paper develops a method for detecting optical binary stars based on the use of astrometric catalogs in combination with machine learning (ML) methods. A computational experiment was carried out on the example of the HIPPARCOS mission catalog and the Pan-STARRS (PS1) catalog by applying the suggested method. It has shown that the reliability of predicting a stellar binarity reaches 90-95\%. We note the prospects and effectiveness of creating a proprietary research platform --- Cognotron.
\end{abstract}



\section{Introduction} \label{sec:intro}

The purpose of this work is to increase the accuracy of estimating the ratio of the number of binaries over single stars by applying artificial intelligence (AI) methods to the classical astronomical techniques.

Binary stars form dynamic systems that rotate under the gravitational attraction around a common center of masses. Binary stars are subdivided into visual binary stars, spectral binary stars, and eclipsed variable stars \cite[pp. 17-21]{Hilditch2001}. Methods of their detection are, therefore, divided into astrometric, spectral, and photometric. 
The development of classical methods of astronomical measurements calls for increasingly complex models, which may include more than ten free parameters per object. As a result, the proportion of discovered stars demonstrating non-linear motions is constantly increasing and, accordingly, the risk of an erroneous estimation of the ratio of binary and single stars is growing. In addition, the classical methods of processing astrometric measurements suffer from the ``curse of dimensionality'' when the computational costs to determine the nonlinear and free parameters grows exponentially with respect to their number.

In this paper, it is proposed to supplement the classical methods of estimating the aforementioned ratio with ML methods. The latter allow to work with a set of poorly defined parameters and reduce the exponential growth of the volume of calculations, arising from their number, to polynomial. However, modern ML methods, as it is well known, do not always enable us to get an explanation of the results obtained. For the purpose of this paper, this is not an obstacle, since the verification of the results derived from the data by ML, as in the classical case, is carried out based on direct observations.

Applications of ML have an appealing potential, which is constantly increasing, primarily due to the development of ML methods and analytical methods of processing big data, as well as the growth of computer power. New versions of ML are being introduced, and the computing capabilities for their implementation are increasing. The development of methods is heading in the direction of providing opportunities to self-study, adapt to the dynamics of the external environment, solve interdisciplinary tasks, plan, analyze, give explanations, consider the subjective, non-local and wave aspects of the behavior of the research objects \cite{Wang2018,Raikov2021}. 

As an example of a successful application of ML in astronomy, one can mention the work \cite{becker2020}, presenting a classifier of astronomical events in real time for Automatic Learning System for the Rapid Classification of Events (ALeRCE). The article \cite{CarrascoDavis2021} 
 discusses photometric and spectroscopic observations of rapidly variable sources formed after the explosion of an astronomical object. To classify signals from the nuclei of galaxies, supernovae, asteroids, etc. Convolutional Neural Network (CNN) is used. At the same time, metadata is added to astronomical images in the form of a priori known functions and indicators, which helps achieve a high level of accuracy ($\approx94$ \%).
There are two types of classification methods --- based on a template and a light curve. The first is able to distinguish a richer taxonomy of events, the second uses only the first event warning. The classification of events by templates is based on the use of CNN \cite{CarrascoDavis2021}. CNN input requires images and metadata about the properties of objects from various catalogs. The standard shape of the event template is $63\times63$ pixels. The list of alert metadata includes about 15 indicators.

The authors of \cite{ALeRCE} proposed a template-based classification method for distinguishing five different classes of events within the framework of ALeRCE, the detection and alert of supernova explosions (SNE), using prediction and distinguish between SNE, as well as other complex classes of events. Template classification is necessary for morphological differentiation of galactic nuclei, SNEs, stars, asteroids and false warnings. It uses rotational invariance of images. CNN is trained using entropy data and additional information that experts could use to evaluate candidates. Modern astronomical instruments are able to assess the level of chaos caused by the explosion of objects \cite{Reyes2018}, estimate the size of a companion star \cite{Jiang2017}, recognize, annotate and classify big data obtained from survey telescopes.

The article is structured as follows. First, the HIPPARCOS catalog is discussed, as well as an overview of the labor of the detection of binaries in this catalog. At the same time, the general principles of allocation of unresolved binary stars in the astrometric survey are discussed. Following this, a review of a calculation experiment using machine learning and ML methods in astronomical research is described. At the same time, considering the features of ML, significant and control signs of duality are distinguished, these signs are selected classifying singular and binary stars. The stability of the constructed classification to changes of the observational selection of the training sample is checked. In conclusion, an estimate of the proportion of binary stars to single stars in the catalog is given.

\section{Ground-based observations and cataloging of binary stars} \label{sec:bin-catalogs}

The idea of the existence of physical binary and multiple star systems in the Universe was first considered by John Michell \cite{michell1767}. He applied then new statistical methods to the study of stars and demonstrated that many more stars occur in pairs or groups than a random distribution can explain. For the Pleiades cluster, Mitchell estimated that the probability of such a close group of stars is $5\cdot10^{-5}$. He concluded that stars in such binary or multiple star systems can attract each other, which is the first proof of the very fact of the existence of binary stars and star clusters. His work on binary stars may have influenced William Herschel's research on the same topic, which took shape in the first catalog of binary stars \cite{hersh1785}.

\subsection{Modern astrometric, photometric, and spectral methods for determining the binarity of stars} \label{subsec:bin-methods}
Identifying objects with a small angular distance as binary stars was initially proposed (the so-called optical double stars, incredibly close to each other in the sky). Later it was supplemented by other methods, primary astrometric (see e.g. \cite{makkap2005}). Long-term observations of optical and astrometric binaries make it possible to determine the orbits of the components in some cases. Currently we are talking about hundreds of such objects \cite{soder1999, hart2001}. This is the only direct method of determining the physical mass of stars.

Astrometric methods are good for investigating sufficiently wide star pairs. For closer components, which are unresolved, astrophysical methods (photometrical or spectroscopic) are more effective.

In most cases, the duality of an object is revealed using criteria such as the Rayleigh criterion \cite{arenou2005,griffin1986}. Optical binary stars were discovered by applying Rayleigh criterion: if the distance between the components exceeds the half-width of the so-called point spread function (PSF), according to earlier methods, one considers the star to be optically binary. It is essential that for more than a hundred years the accuracy of determining the astrometric parameters of stars has been much better than PSF. This is related to the development of astrometric methods for detecting binary stars based on the features of their proper motions. Our work is also aimed at clarifying the criteria for detecting the duality of astrometric binary stars, which would be orders of magnitude more sensitive than the Rayleigh criterion.

\subsection{Binary star catalogs, binarity in stellar surveys} \label{subsec:bin-catalogs}

Catalogs of binary stars have been published since the end of the XVIII century \cite{hersh1785}.
The development of the situation is shown in the table 1 for publications \cite{burnham1906, aitken1932, IDS1963, Lipaeva2014, wds1984, wds1997, wds2001, wds2021, CCDM1994, TDS2002}. It is easy to see that the growth of number of discovered binary systems is quite moderate: at the beginning of the XX century, astronomers used visual catalogs containing a total of about a million stars \cite{SD1886,CpD1896,CpD1897,CpD1900,BD1903}, in the middle of the century, a photographic Carte du Ciel with about 4.5 million stars \cite{Eichhorn1957, CdC1972,CdC1983} became available, the development of space astronomy required catalogs of tens of millions of objects \cite{GSC1990a,GSC1990b,GSC1990c,GSC12_2001}, and in the XXI century, electronic versions of photographic catalogs with a volume of about a billion stars were introduced to the scientific community \cite{GSC2_2008,USNOB_2003}. During this period, the volume of double star catalogs has grown only a few times.

\begin{table}
	\caption{Catalogs of double stars}\label{tab:catadouble}
	\begin{center}
	\begin{tabular}{|l|r|r|r|p{2.7in}|}
	\hline
	ID & Year & \multicolumn{2}{c|}{Total number of} & References\\
	~ &  ~    &\multicolumn{1}{p{1in}|}{multiple systems} &\multicolumn{1}{p{1in}|}{individual components} &  \\
	\hline
	Hershel& 1785& 434& ~ & \cite{hersh1785}	\\
	BDS & 1906& 13665 & ~ & \cite{burnham1906}\\
	ADS & 1932 & 17180 & ~ &\cite{aitken1932}\\
	IDS & 1961 & 56572 (29965) & 69819 (36861) & \cite{IDS1963,Lipaeva2014}\\
	WDS & 1994&  73610 & 154333 & \cite{wds1984,wds1997,wds2001,wds2021}\\
	CCDM &1994,2002  &34031& 74861&\cite{CCDM1994}\\
	Tycho-3&2001 &32631 &103259 & \cite{TDS2002} \\
	\hline
	\end{tabular}
	\end{center}
\end{table}

The above situation is typical for survey catalogs limited to the weakest observed stellar magnitude. Most of the stars in such a catalog \cite{kharchenko_2001} are slightly brighter than the detection limit and identifying duality in this case is problematic.

A survey of the star catalogs, compiled according to a somewhat different principle: identification of all objects within a given volume of space \cite{Gliese1991}, significantly changes the statistics \cite{imf_2010, Duquennoy1991a,Duquennoy1991b}. 
Dimmer objects of late spectral classes are beginning to prevail in the samples of the nearest to the Sun stars, and the proportion of stars with signs of duality is approaching 50\%.

\subsection{Theoretical models of the emergence of multiple stars, the percentage of binarity among the stars in the vicinity of the Solar System} \label{subsec:bin-theory}

There is no common opinion in the research of the mods of the collapse of protostellar clouds yet.  For the appearance of single stars, theoretical models are well-regarded, and a sufficiently convincing initial mass function is obtained, which then gives reasonable results in further population calculations. So, according to \cite{kroupa2002}, there are following equations for different masses of stars:

\begin{equation*}
\xi (M)={\begin{cases}k_{0}\left({\frac {M}{m_{0}}}\right)^{-\alpha_{0}}&,\quad m_{0}<M\leqslant m_{1}\\k_{1}\left({\frac {M}{m_{1}}}\right)^{-\alpha_{1}}&,\quad m_{1}<M\leqslant m_{2}\\k_{2}\left({\frac {M}{m_{2}}}\right)^{-\alpha_{2}}&,\quad m_{2}<M\end{cases}} 
\end{equation*}	

\noindent where $m_0 = 0.01m_\odot$, $m_1 = 0.08m_\odot$, $m_2 = 0.5m_\odot$, $\alpha_0 = 0.3$, $\alpha_1 = 1.3$, $\alpha_2 = 2.3$. The last equation is the initial mass function of \cite{salpeter1955}.

There is no generally accepted model of binary star formation yet. There are several theoretical models of the collapse of protostellar clouds with an initial rotation. In such models gas compression must lead to the formation of a toroidal structure. This structure then decays into separate protostars, which form a multiple star system (in the simplest case, a binary one). Taking into account various physical mechanisms (for example, the degree of influence of the magnetic field) leads to significantly different results, none of which is fully confirmed by observations.

Our study of binary stars statistics is carried out under conditions \cite{Gliese1991} of extremely high probability of duality of objects (about 50\%)and the absence of a generally accepted astophysical theory.

\section{HIPPARCOS main catalog} \label{sec:hipparcos}

The HIPPARCOS spacecraft became the first specialized astrometric satellite. High-precision optical measurements require a long-focus instrument, which therefore has a small field of view. 

\subsection{Observations technique and the structure of the catalog. Sources of information and data on binarity} \label{subsec:hip-technic}

The task of carrying out high-precision measurements throughout the entire celestial sphere determined the instrument design with two fields of view with a diameter of about $0.9^\circ$ each, spaced from each other by $\approx58^\circ$. The device rotated with a period of 120 minutes around an axis perpendicular to the plane in which the entrance pupils lie \cite{hipp1989a}, \nocite{hipp1989b} and the axis itself (according to the observation plan) slowly precessed along the cone of $43^\circ$ around the direction to the Sun. All these movements combined led to a uniform coverage of the observations of the celestial sphere and, on the other hand, did not allow the satellite's solar panels to deviate significantly from the direction of the Sun and lose power.

To increase the stability of the measurements, it wasn't just the passage of a star in the field of view of an electric vacuum device that was being recorded, but its passage through a lattice of 2660 slits. The recorded coordinates thus ended up being one-dimensional, they were tied to a large circle, that remained the same for several revolutions of the satellite. During the processing stage, the parameters of the circles were linked to each other for the entire celestial sphere, then the spherical coordinates of individual stars, their parallaxes and proper motions were calculated \cite{hipp1989c}.

The results of the satellite observation \cite{hip1997a} have shown the accuracy of coordinates better than of ground observations by about 100 times. During the experiment, this made it possible to determine the coordinates alongside parallaxes and proper motions with high accuracy of about 1 msec of arc. That is, an experiment of 3.5 years yielded a result comparable to the astrometric activity of an entire century and in some cases (in terms of high-precision parallaxes), even surpassing the results. We have to note that, strictly speaking, this is correct only of the objects of the HIPPARCOS program, with respect to a posteriori confirmation of the accepted source model.

The duality of objects in the HIPPARCOS main HIP catalog was initially established by their belonging to the Catalog of Components of Double and Multiple Stars (CCDM) \cite{CCDM1994}.  
At the same time, new optical binaries reliably detected during the observations, separated by the Rayleigh-type criterion, were added to this catalog (about 5\% of its volume).

\subsection{Further work on the detection of double stars in HIP}\label{subsec:hip-detection}

After the publication of the main HIPPARCOS catalog and the additional Tycho catalog to the original program, their in-depth research began. The latest catalog was obtained based on observations not of the main HIPPARCOS spacecraft photodetector, but of the signals from service stellar sensors, initially intended to monitor the appearance of an object that is a part of the observation program in the field of view immediately before the start of the main measurements. This additional material allows us to get the coordinates of about a million stars with an accuracy of an order of one to one and a half magnitude worse than in the main HIP catalog.

Considerable efforts of researchers were aimed at improving the accuracy of the Tycho catalog's proper motions and increasing its size based on satellite observations records and the use of Carte du Ciel data, so Tycho-2 \cite{Tycho-2} catalog appeared. Later on, this information formed an additional catalog of binary stars Tycho \cite{TDS2002}  and the discovery of new binary stars in the main HIP catalog \cite{makkap2005}.
  We will use the latter list below to test the operation of machine learning algorithms.

\subsection{General principles of detection of unresolved binaries in an astrometric survey} \label{subsec:hip-general}

It is apparent, that many different criteria for detecting the duality of an object by measuring any one parameter, such as the ellipsoid of the visible image, the deviation of proper motion from a straight line, anomalous photometry, bifurcation of spectral lines ? have already been used, generally speaking, by the authors of observations or their closest followers and cannot yield anything new. The general principle of detecting duality in this case is approximately the same: the Rayleigh criterion in the latter case or the ellipticity of the image significant in comparison with the point spread function (PSF) in the first case, are similar to each other and do not make it possible to detect duality if the separation (or anomalous proper motion) of the components is less than the errors, the width of the spectral line or the same PSF. 

On the other hand, it is obvious that the optical separation of the components of the assumed binary at distances greater than the errors of the coordinate measurements should somehow manifest itself in the results of the observations. Our work consists of verifying the fact that for unresolved, as well as for resolved, binary stars, anomalies in the errors of the measured values will appear.

The proposed method relies on using the widest possible set of data for each object. The next section shows how, based on a training sample, various artificial intelligence algorithms are trained, significant signs of duality and signs of negligible significance are highlighted. An additional check is also performed on the sampling effect, the influence of the chosen machine learning method, the influence of the observational selection in the training sample and the dependence on the imperfection of the reduction procedure of the main catalog. 
In addition, some of those parameters of stars that should not depend on duality are to be analyzed for the purpose to check the algorithms used.

\section{Application of ML methods for detection of binary star systems}\label{sec:ML}

In this work we present the results of applying modern ML methods and AI systems to the data of HIP catalog to increase the quality of the data received via the astrometric satellite. More specifically, we used neural networks and decision tree-based models.

\subsection{Feature extraction}\label{subsec:ML-features}

The HIP catalog contains many characteristics of stars from the main list (there are 77 data fields in the catalog \cite{hip1997a}). From them, we excluded fields like references to other catalogs or data sources. We also excluded astronomical coordinates, because on the one hand, the catalog covers only a small volume of the Galaxy in a close neighborhood of the Sun, where the structure of the Galaxy is not prominent. On the other hand, exact values of star coordinates are enough to uniquely identify a star, so the ML models will just remember which stars are binary and which are not, without trying to extract dependencies from other fields.

The spectral class field of a star is presented in the HIP Catalog as a text field, which prevents its direct use as an input to ML models. To use information from this field, the text format was converted to a synthetic (manually created) feature vector which describes the spectral data in a format, suitable for ML models. A complete set of features of this vector is shown in the Appendix A. 
Thus, the catalog data was mapped in a space of 73 features, including 34 numeric values taken directly from the HIP catalog and 39 features from the spectral classification fields. The table showing all catalog fields used in the analysis is shown in the Appendix B. 

Non-empty value of CCDM field of HIP catalog was used as a target feature.

\subsection{Data analysis models}\label{subsec:ML-models}

In the AI research platform we have experimented with two kinds of models to process data: ensemble of neural networks and ensemble of XGBoost (eXtreme Gradient Boostring \cite{XGBoost2016}) decision-tree trained models.

Neural network ensembles were built from feed-forward networks composing of two fully-connected hidden layers of 200 and 100 neurons with ReLU (Rectified Linear Unit) activation function, followed by an output layer of a single neuron with a sigmoid activation function. Before feeding data to the neural network, it was passed through a normalization layer. Networks was trained using Adam (Adaptive Moment Estimation) \cite{Adam2015} algorithm, optimizing for a binary cross-entropy metric. Neural networks were implemented and trained using Keras \cite{Keras2015} library, included in Tensorflow deep learning framework  \cite{tensorflow2015}.

To create an ensemble of neural networks, we split the original HIP catalog dataset into 25 random subsets, while preserving the proportion of target binarity feature in each subset. This was done using StratifiedShuffleSplit function from sklearn library \cite{sklearn2011}.
Then, each single subset was used as a training datum for a neural network, while the remaining 24 subsets where jointly used as a test set. Those 25 networks were grouped into an ensemble, which was used to calculate the final output value for all stars in the HIP Catalog.

An ensemble of decision-tree models was trained using the same technique. Base models were created using XGBClassifier implementation of the extreme gradient boosting algorithm from XGBoost library \cite{XGBoost2016}.
The following parameters were used for each of 25 classifiers: {\tt learning\_rate=0.01}; {\tt n\_estimators=1811}; {\tt max\_depth=6}; {\tt min\_child\_weight=4}; {\tt gamma=0.4}; {\tt subsample=0.9}; {\tt colsample\_bytree=0.8}; {\tt objective= 'binary:logistic'}; {\tt nthread=4}; {\tt scale\_pos\_weight=1}; {\tt seed=27}. Parameters, essential for learning were found using a cross-validation method.

Since there is reason to believe that not all multiple systems are marked in the HIPPARCOS catalog, we propose to use an approach that is often used in data analysis to identify labeling errors, to identify candidates for binary systems: train a classifier on an initial (incomplete in terms of labeling) data set and then consider objects with the maximum error as candidates for incorrect labeling \cite{Brodley1999,Zhu2003,Angelova2005,Huang_2019}.

In the context of our task this means that we are looking for stars, which are not marked as binary in HIPPARCOS catalog, but, at the same time, show a high probability of being binary based on the results of the ML models, trained on the datasets from the catalog. For such an object, it holds that, despite the fact that it was used in the training process as ``not a binary star'', i.e.  the ML model was instructed that the probability of its binarity equals 0, the ML algorithm, nevertheless, insists on its duality, based on the patterns it derives from the data.

\subsection{Importance of features required for classifying a star as binary}\label{subsec:ML-ranging}

\begin{figure}[ht!]
\plotone{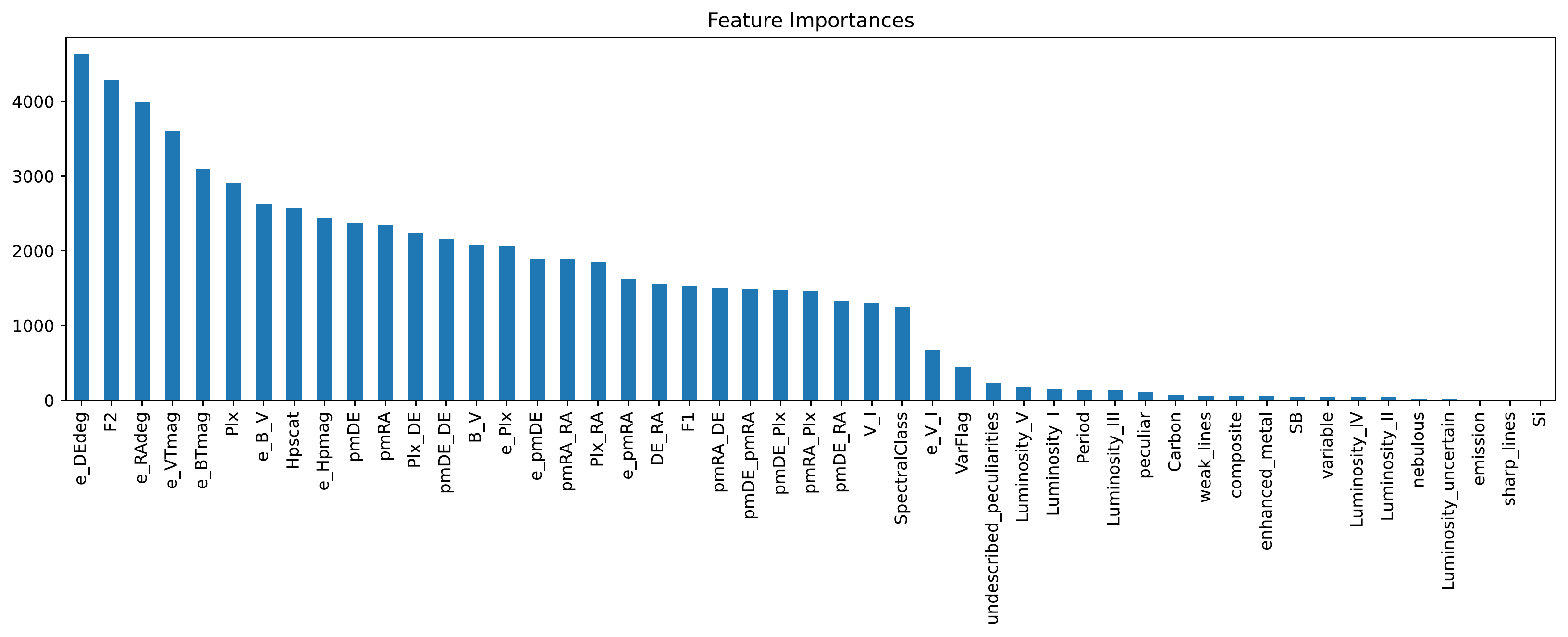}
\caption{Features from the fields of HIPPARCOS catalog ranked by importance to produce the probability of duality of a star
	\label{fig:1-features}}
\end{figure}

XGBoost library provides means to access various statistics of a model, including feature importance's for trained classifier. Fig. \ref{fig:1-features} shows that various statistical parameters from the catalog are significant, while, for example, the spectral parameters are not. It means that ML algorithms confirmed the well-known from the catalogs of double and multiple stars result, that duality of a star is only loosely correlated with the spectral characteristics of the pair. Thus, we can conclude that training dataset used is consistent in this regard.

Importance score shown on x-axis of Figure \ref{fig:2-robustness} represents the number of times the feature was used as a decision variable in trees, averaged over ensemble members. Our calculation experiments show that this feature is consistent across ensemble members and resilient to changes at random stages of algorithm training, like a split of the ensemble members into individual training groups, seed parameter of XGBClassifier.

\subsection{Robustness of the classification algorithm}\label{subsec:ML-robustness}

The training set for ML algorithms consists a subcategory of catalogs stars \cite{hip1997a} 
assigned a non-empty value of CCDM \cite{CCDM1994} field, i.e. the star is recognized as a double. The very procedure for establishing binarity based on the anomalous proximity of neighboring stars (see above) is subject to a very strong observational selection: binarity according to catalogs \cite{CCDM1994, wds2021} for bright stars (of which there are few) turns out to be much more probable than for more numerous faint stars. In order to test the influence of this effect on the work of ML algorithms, an independent training was carried out, during which photometric values from model input were excluded. The specific parameters that were used at this stage are given in the table in the Appendix B. The star color indices and photometric errors were retained as input parameters: the former --- to control the adequacy of the classification algorithm, the latter --- as one of the indicators of the possible duality of the object.

\begin{figure}[ht!]
	\plotone{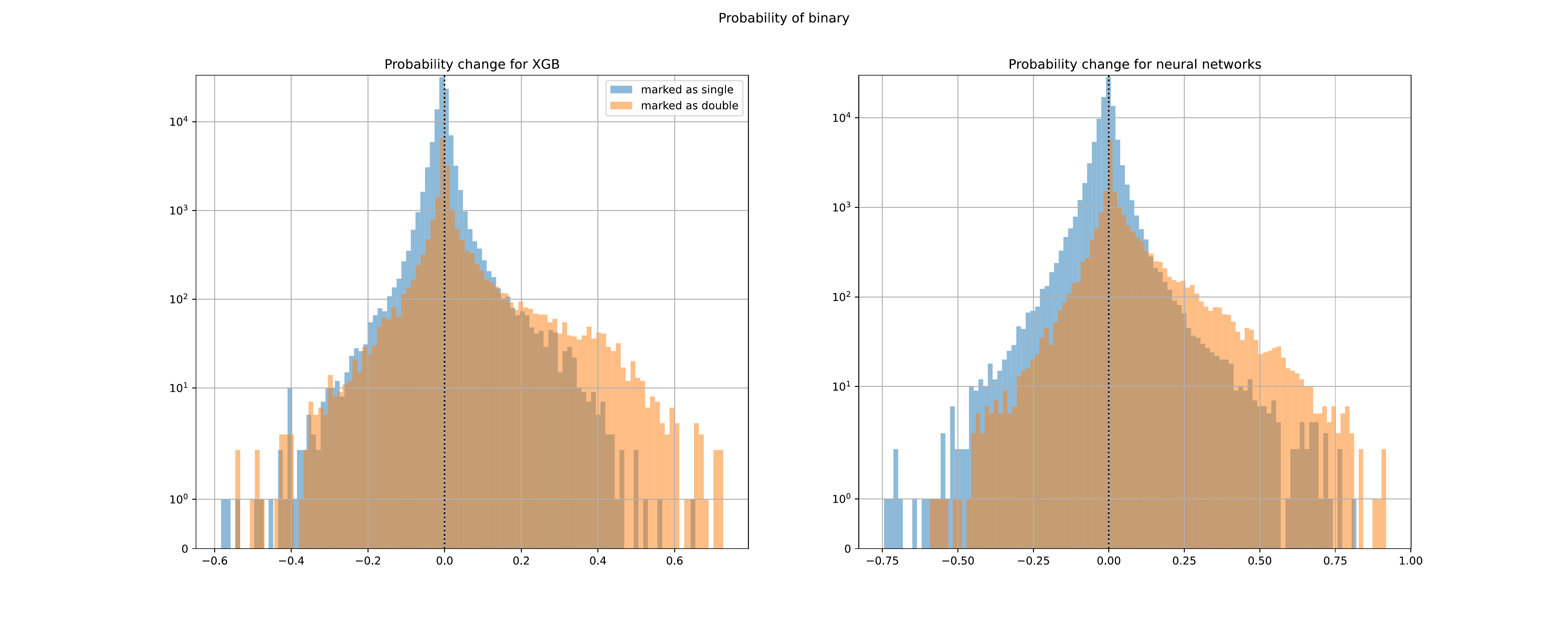}
	\caption{The changes of probabilities of classifying a star as a singular or binary by different ML algorithms
		\label{fig:2-robustness}}
\end{figure}

The results of the work of the HIPPARCOS consortium published in \cite{hip1997a}, included the solution of a system of nonlinear reduction equations for determining the kinematic parameters of stars. The system of equations was solved by the iteration method. One iteration in 1995-1996 took about half a year of calculations, and it was this iteration that was subsequently published. After 10 years, one of the members of the consortium repeated the processing of the observational material that had been preserved, somewhat improving the reduction scheme and achieving convergence of the iterations (at that time, one iteration took about a week of computing time). The resulting new HIPPARCOS reduction was published \cite{hip_new2007a, hip_new2007b} and showed slightly better accuracy, especially for bright stars. It has not been accepted as a coordinate standard, but is actively being used in scientific research.

To test the approach proposed in this paper, this new reduction is important, since it allows one to check the stability of ML algorithms to the non-Gaussian nature of the errors of the parameters being analyzed. Since the convergence of the solution of the system of nonlinear equations was not achieved in the HIP catalog (it was only possible to check the absence of noticeable divergence), the errors of the quantities being determined will inevitably have a non-zero mean, which indicates that they will be non-Gaussian.

The training of ML models was carried out using data from the original HIP catalog and data from its new reduction together. A specific list of fields taken from the new reduction \cite{hip_new2007b} is given in the table in the Appendix C. Also, two Boolean fields were added, indicating the presence of a 7- and 9-component solution (the presence of a star in the \textbf{hip7p.dat} and \textbf{hip9p.dat} tables from \cite{hip_new2007b}). The specific values of these solutions were not used, since they are available only for a small part of the catalog stars, and it is unlikely that they will be effectively used by the ML models in this regard.

The experiment with machine learning based on the material \cite{hip_new2007b}
showed stability of the values of the main statistical characteristics in the catalog for the purpose of classifying objects as binary stars (Fig. \ref{fig:2-robustness}). The resilience of the results to the sampling effect was also tested.

\subsection{Prediction of duality probability of HIP catalog stars by ML models}\label{subsec:ML-predict}

\begin{figure}[ht!]
	\plotone{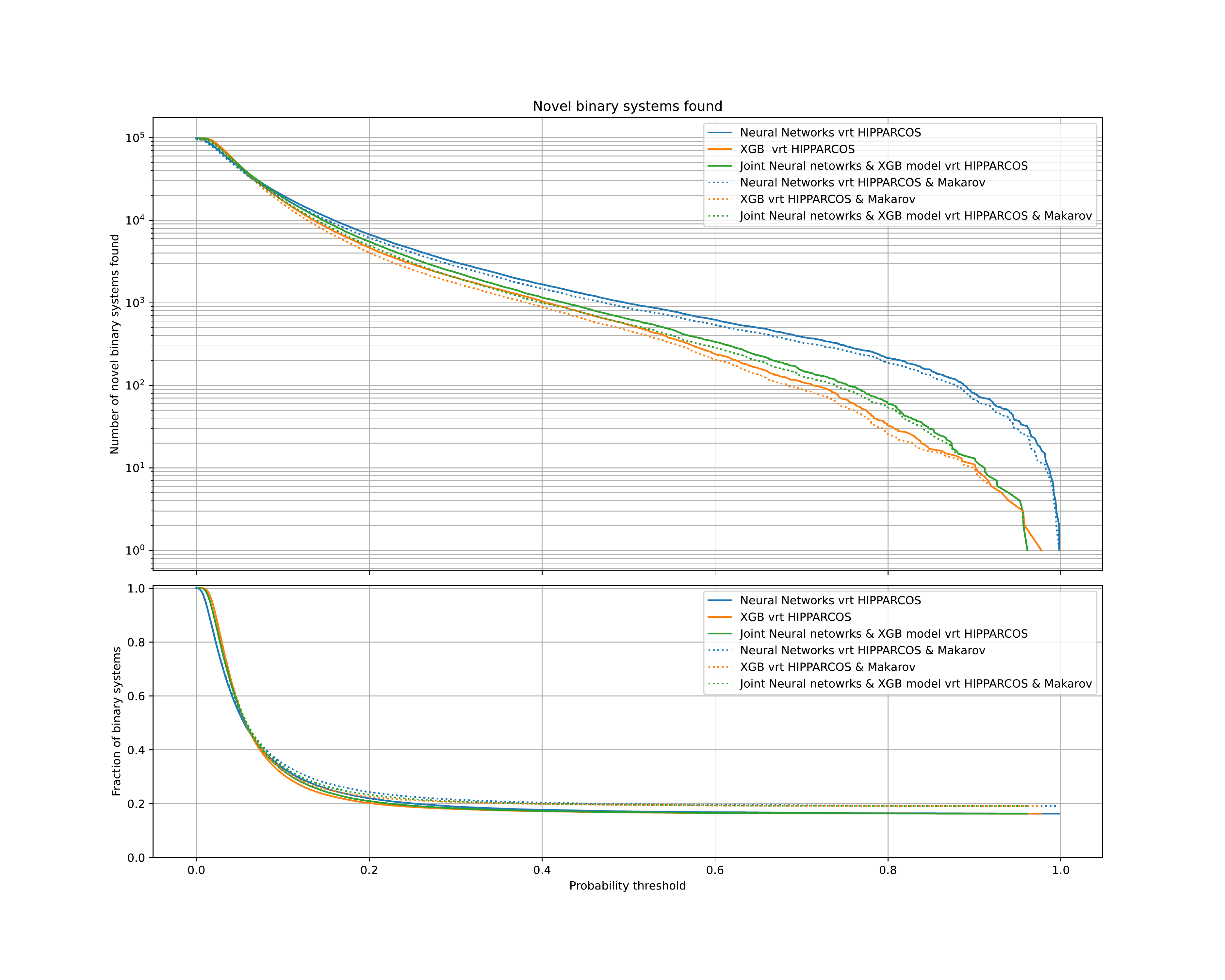}
	\caption{Number of candidates to double systems as a function of threshold on model probability
		\label{fig:3-prediction}}
\end{figure}

Since the output of the proposed models is the probability of duality for each of the stars of the HIPPARCOS catalog, it is natural to consider stars, for which probability exceeds a certain threshold as candidates for binary stars. Figure \ref{fig:3-prediction} shows how many new candidates for binary systems can be identified using the proposed models for different threshold values. The solid lines correspond to binary system candidates identified in comparison with the labeling of the original HIP catalog, and the dotted lines correspond to those found in work \cite{makkap2005}. The graph at the bottom of the figure shows what the percentage of multiple systems in the catalog will be at different values of the probability threshold.

As outputs of two models --- ensemble of neural networks and ensemble of decision trees differ from each other, it is a natural next step to group these models together into one joint ensemble including models of both types. In Fig. \ref{fig:3-prediction}, along with the results of the ensemble of neural networks and the ensemble of decision trees, the results of the combined model are also presented.

The data in Fig. \ref{fig:3-prediction} correspond to the results obtained when training models using the data from both the original HIP catalog \cite{hip1997b} and the new reduction \cite{hip_new2007b}, using a set of variables that limits the effect of observational selection. The methodology used in this work does not imply the identification of all binary star candidates in one run. Solving the problem of data analysis under conditions of partially incorrect labeling \cite{Brodley1999,Zhu2003,Angelova2005,Huang_2019} implies an iterative process, during which after identifying the most probable labeling errors (in our case, previously unidentified double stars), it is necessary to check the corresponding objects, correct the labeling and repeat training. This cycle can be repeated several times until a satisfactory result is achieved. This article presents the first iteration of this process.

\subsection{Comparison of ML methods' results with other publications}\label{subsec:ML-compare}

The comparison was carried out on the most extensive work on the identification of additional double stars in the HIP catalog --- the paper by Makarov and Kaplan \cite{makkap2005}, where the astrometric method was used, with additional information from the Tycho-2 catalog \cite{Tycho-2}.

\begin{figure}[ht!]
	\plotone{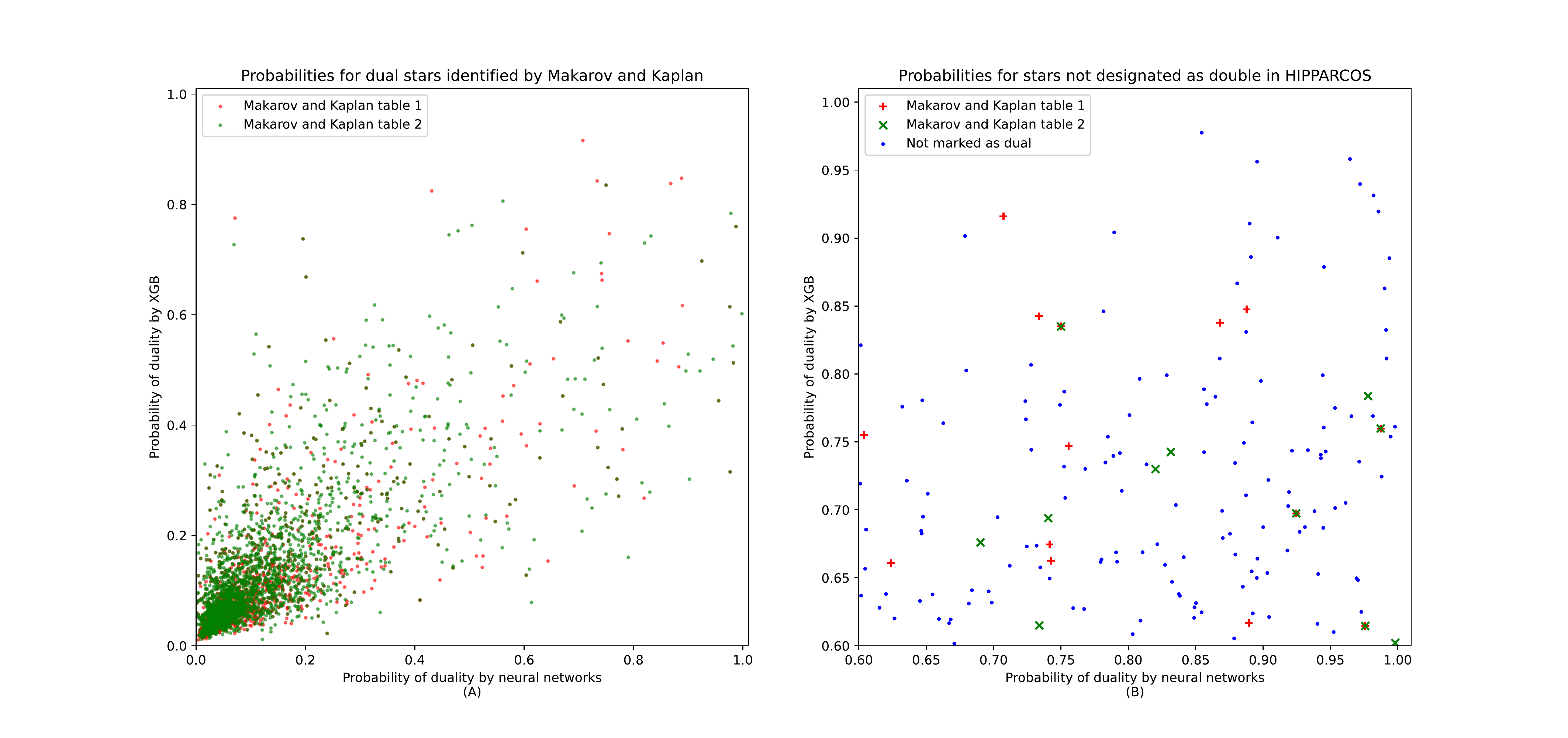}
	\caption{Comparison of ML algorithms' results with Makarov and Kaplan 
		\label{fig:4-compare}}
\end{figure}

Figure \ref{fig:4-compare}A shows the probabilities computed by our ML models for stars identified as double in \cite{makkap2005}. The axes of the graph correspond to probabilities of duality calculated by the ensemble of neural networks and the ensemble of tree classifiers. 

In Figure \ref{fig:4-compare}B there are also stars that are not designated as double either in \cite{makkap2005} or in the original HIPPARCOS catalog along with the stars identified in \cite{makkap2005}, with a binarity probability estimates of $> 60$\% from the ML models.

Analyzing graph \ref{fig:4-compare}(A), one can see that most of the multiple stars identified in \cite{makkap2005} received low probabilities of duality when assessed using the ML system. On the other hand, plot \ref{fig:4-compare}(B) shows that most of the stars proposed as candidates for duality by the ML models were not identified in \cite{makkap2005}. Thus, we can conclude that the patterns identified by machine learning models and the corresponding candidate stars differ from the results of \cite{makkap2005}.

In this regard, it may be of interest to include the binary stars identified in \cite{makkap2005} in the training set, to train and study the resulting models.

\subsection{Observation-based model verification} \label{subsec:observ}

To verify the obtained results of applying ML to the objects of the HIPPARCOS catalog, the lists of the most probable double stars were compared to the double stars from the Pan-STARRS catalog of objects \cite{PS1}. The search for neighbors was carried out in the vicinity of the HIPPARCOS star with a radius of $5''$. This is the size of the working area of the photodetector that took the measurements.

\begin{table}
	\caption{Identification of double star candidates in PS1}
	\label{table:panstarrs}
	\begin{center}
	\begin{tabular}{|l|r|r|r|r|}
		\hline
	selection criteria & number of objects & \multicolumn{1}{p{1.2in}|}{multiplicity found (components found)} & \multicolumn{1}{p{1.2in}|}{components not found} & star not found\footnote{the catalog PS1 was obtained from the observations of the telescope installed in the Hawaiian Islands, part of the sky is not available for observation}
		\\
		\hline
		NN, $p>0.8$ & 214 & 142 (895*) & 10\footnote{2 spectroscopic binaries, 3 stars with large proper motion, 1 against the background of the galaxy} & 72
		\\
		\hline
		XGB, $p>0.7$ & 109 & 69 (430*) & 5\footnote{3 spectroscopic binaries } & 50
		\\
		\hline
	\end{tabular}
	\end{center}
\end{table}

Thus, the developed mechanism gives a 90-95\% probability of correct prediction of duality (recall that a priori the probability of duality of a randomly selected star is about 50\%).

\section{Conclusions} \label{sec:conclusions}

The methodological approach developed in the Cognotron research platform and presented in the article and the experiments performed show that the use of machine learning methods on the data of the HIPPARCOS catalog makes it possible to extract additional information and identify double and multiple star systems that could not be previously detected by classical methods. This is the result of a discovery of complex relationships between astrometric and photometric characteristics and the errors of these characteristics by machine learning methods. Classical methods are based on an analysis of isolated characteristics, or small groups of characteristics, and are limited by the accuracy of their measurements. The combination of a larger number of characteristics and their errors in the analysis, which is achieved by using machine learning methods, is equivalent to using a larger number of accumulated light quanta during long-term observation, which makes it possible to increase the accuracy of detecting binary stars.

The disadvantage of the proposed method is that it does not allow introducing a strict criterion for the duality of stars. In classical methods, criteria of this kind are formulated on the basis of known physical laws prior to the analysis. However, it is not theoretically possible to formulate such a criterion that would describe the relationship between several dozen characteristics. Machine learning methods in this case rely on an automated extraction of dependencies from labeled data. But, in the case in picture, there is no correctly labeled training sample, and any patterns are extracted from data in which a significant part of binary stars are not labeled as such.

An immediate consequence of this situation is that the output values of the data analysis models proposed in this paper cannot be considered as probabilities of stellar duality. Since the proportion of binary stars in the training sample is underestimated, it should be expected that the output values will also be underestimated compared to the true probabilities.

The procedure for detecting binary stars is also becoming more complicated. To fully use the potential of the method proposed in this paper, it is necessary to implement an iterative procedure, during which the binary star candidates proposed by the ML models will need to be independently verified; the labeling of the training data will change based on this verification, new models will be trained that will offer the next set of candidates and so on.

This article, in fact, presents the results of one such iteration. It is important that this work demonstrates that during such iteration, it is indeed possible to select objects with a high probability of duality (90\% of the objects were confirmed according to the PS1 catalog). It means that:
\begin{itemize}
	\item 
in the HIPPARCOS data, there are indeed significant dependencies indicating the duality of stellar systems;
	\item 
existing machine learning methods allow such dependencies to be detected and used to identify dual stars that were not detected by classical methods.
\end{itemize}
Significant features identified by ML algorithms for the dual star classification are of high interest as well. In the process of training and verifying ML models, it turned out that machine learning algorithms quite reliably identify a group of significant features associated with the statistical characteristics of the observed values. It is also shown that the identification of duality only loosely depends on the spectral characteristics of the pair.
The methods turned out to be resilient to observational selection in the training set itself. In addition, the parameters of objects in the HIPPARCOS catalog, which, according to the studies of other authors, are not related to the multiplicity/duality of stars (for example, various spectral characteristics), showed low significance, which was additional evidence of the effectiveness of this method.

The application of the digital ML methods-based approach proposed in this paper to data and catalogs of other missions (for example, GAIA, \cite{GAIA_EDR3_2021}) is also possible and promising, but so far seems premature. Combining data across multiple missions is of theoretical interest, but may be accompanied by difficulties due to the difference in the values obtained by different instruments and due to different observation schemes. In addition, their statistical characteristics will also differ, which may introduce difficulties for ML algorithms. An additional complication arises from the difference in the operating ranges of stellar magnitudes, although an intersection does take place. Such a combination is also complicated by the fact that, according to pre-flight plans, the publication of data on relatively complex objects will be carried out at the final stages of the experiment.

Despite the difficulties and obstacles that arise, the ML approach proposed in this paper is more accurate in its measurements and can help extract new knowledge and stimulate the generation of new ideas, compared to the classical approach. A specific AI platform named ``Cognotron'' was used in our study. 

\newpage	
\appendix

\section{} \label{sec:appA} 

\startlongtable

\begin{deluxetable*}{llp{2in}p{1in}}
	\tablecaption{Spectrum Description fields\label{tab:spec}}
	\tabletypesize{\footnotesize}
	\tablewidth{0pt}
	\tablehead{
		\colhead{Field} &
		\colhead{Range} &
		\colhead{Description} &
		\colhead{Notation in the field SpType}
	}
	\decimalcolnumbers
	\startdata
		SpectralClass &
		0...8 &
		Spectral class &
		O, B, A, F, G, K, M, L, T, C, S, SC, WN, WC, WO, WR, R, N, DA, DB, DC, DO, DZ, DQ, DG, DF, CN\\
		\hline
		Luminosity\_I &
		Boolean &
		Luminosity Class I &
		I, Ia, Iab, Ib\\
		\hline
		Luminosity\_II &
		Boolean &
		Luminosity Class II &
		II, IIa, IIb\\
		\hline
		Luminosity\_III &
		Boolean &
		Luminosity Class III &
		III, IIIa, IIIb\\
		\hline
		Luminosity\_IV &
		Boolean &
		Luminosity Class IV &
		IV, IVa\\
		\hline
		Luminosity\_V &
		Boolean &
		Luminosity Class V &
		V, Va, Vb\\\hline
		Luminosity\_VI &
		Boolean &
		Luminosity Class VI &
		VI\\\hline
		Luminosity\_uncertain &
		Boolean &
		Luminosity class is not precisely defined &
		:\\\hline
		subdwarf &
		Boolean &
		Subdwarf &
		sd\\\hline
		WhiteDwarf &
		Boolean &
		White dwarf &
		DA, DB, DC, DO, DZ, DQ, DG, DF\\\hline
		W &
		Boolean &
		Wolf-Rayet star &
		WN\\\hline
		Carbon &
		Boolean &
		C-type star (Carbon star) &
		C\\\hline
		S &
		Boolean &
		S-type star
		(Zirconium star) &
		SC\\\hline
		R &
		Boolean &
		Spectral class R &
		R\\\hline
		N &
		Boolean &
		Spectral class N &
		N\\\hline
		MN &
		Boolean &
		~ &
		MN\\\hline
		nebulous &
		0, 1, 2 &
		Wide spectrum lines &
		n, nn, n:\\\hline
		enhanced\_metal &
		Boolean &
		Strong Metal Lines &
		m, m:\\\hline
		peculiar &
		Boolean &
		Spectrum Pecularities &
		p, p:, +pec\\\hline
		shell &
		Boolean &
		Shell Star &
		sh, +shell, shell\\\hline
		emission &
		Boolean &
		Emission lines &
		e, e:, eq:, E\\\hline
		weak\_lines &
		Boolean &
		Weak lines &
		w, wk, wl\\\hline
		sharp\_lines &
		Boolean &
		Narrow lines &
		s, ss, s:\\\hline
		variable &
		Boolean &
		Variable spectr &
		v, va, var\\\hline
		weak\_helium &
		Boolean &
		Weak lines of Helium &
		Hewk\\\hline
		NIIandHeIIEmission &
		Boolean &
		N~III emission, absence or weak absorption of He~II  &
		(f)\\\hline
		HeIIabsorbtionNIIemission &
		Boolean &
		strong He II absorption, weak N III emission &
		((f))\\\hline
		composite &
		Boolean &
		A spectrally double star &
		comp\\\hline
		undescribed\_peculiarities &
		Boolean &
		Other features &
		., ..., .., +..., +.., +., +....\\\hline
		SB &
		Boolean &
		Spectroscopic binary &
		SB, SB1, SB:\\\hline
		Sr &
		Boolean &
		Spectral lines of Strontium &
		Sr, Sr:\\\hline
		Cr &
		Boolean &
		Spectral lines of Chromium &
		Cr\\\hline
		Eu &
		Boolean &
		Spectral lines of Europium &
		Eu\\\hline
		Mn &
		Boolean &
		Spectral lines of Manganese &
		Mn\\\hline
		Hg &
		Boolean &
		Spectral lines of Mercury &
		Hg, Hg:\\\hline
		Si &
		Boolean &
		Silicon spectral lines &
		Si\\\hline
		Li &
		Boolean &
		Lithium spectral lines &
		Li\\\hline
		Del &
		Boolean &
		Spectrum like~Delta Delphini &
		delDel, dDel, deltaDel\\\hline
		Nova &
		Boolean &
		Nova Star &
		Nova\\\hline
\enddata
\end{deluxetable*}

\newpage

\section{} \label{sec:appB} 

\startlongtable

\begin{deluxetable*}{llcc}
	\tablecaption{Used parameters from the HIPPARCOS catalog\label{tab:hip}}
	\tabletypesize{\footnotesize}
	\tablewidth{0pt}
	\tablehead{
		\colhead{Name} &
		\colhead{Description in the HIPPARCOS catalog} &
		\colhead{Used in the full set} &
		\colhead{Used in a set}\\
		\colhead{} &
		\colhead{} &
		\colhead{} &
		\colhead{with a restriction of}\\
		\colhead{} &
		\colhead{} &
		\colhead{} &
		\colhead{observational selection}
	}
	\decimalcolnumbers
\startdata
Name 	&	 							&	&	\\\hline
Catalog &	Catalogue (H=Hipparcos)   	&	&	\\\hline
HIP 	&	Identifier (HIP number)   	&	&	\\\hline
Proxy 	&	Proximity flag   			&	&	\\\hline
RAhms 	&	Right ascension in h m s, ICRS (J1991.25)  	&	&	\\\hline
DEdms 	&	Declination in deg ' $''$, ICRS (J1991.25)  &	&	\\\hline
Vmag 	&	Magnitude in Johnson V   					& + &	\\\hline
VarFlag &	Coarse variability flag   					&+ 	&+	\\\hline
r\_Vmag &	Source of magnitude   						&	&	\\\hline
RAdeg 	&	alpha, degrees (ICRS, Epoch=J1991.25)   	&	&	\\\hline
DEdeg 	&	delta, degrees (ICRS, Epoch=J1991.25)   	&	&	\\\hline
AstroRef &	Reference flag for astrometry   			&	&	\\\hline
Plx 	&	Trigonometric parallax   					&+ 	&+	\\\hline
pmRA 	&	Proper motion mu\_alpha*cos(delta), ICRS 	&+ 	&+	\\\hline
pmDE 	&	Proper motion mu\_delta, ICRS   			&+ 	&+	\\\hline
e\_RAdeg &	Standard error in RA*cos(Dedeg) 			&+ 	&	\\\hline
e\_DEdeg &	Standard error in DE   						&+ 	&	\\\hline
e\_Plx 	&	Standard error in Plx   					& + & +\\\hline
e\_pmRA &	Standard error in pmRA   					& + & +\\\hline
e\_pmDE &	Standard error in pmDE   					& + & +\\\hline
DE\_RA 	&	Correlation, DE/RA*cos(delta) 				& + & +\\\hline
Plx\_RA &	Correlation, Plx/RA*cos(delta) 				& + & +\\\hline
Plx\_DE &	Correlation, Plx/DE   						& + & +\\\hline
pmRA\_RA &	Correlation, pmRA/RA*cos(delta) 			& + & +\\\hline
pmRA\_DE &	Correlation, pmRA/DE   						& + & +\\\hline
pmRA\_Plx &	Correlation, pmRA/Plx   					& + & +\\\hline
pmDE\_RA &	Correlation, pmDE/RA*cos(delta) 			& + & +\\\hline
pmDE\_DE &	Correlation, pmDE/DE   						& + & +\\\hline
pmDE\_Plx &	Correlation, pmDE/Plx   					& + & +\\\hline
pmDE\_pmRA &	Correlation, pmDE/pmRA   				& + & +\\\hline
F1 &		Percentage of rejected data   				& + & +\\\hline
F2 &		Goodness-of-fit parameter   				& + & +\\\hline
H31 &		HIP number (repetition)   					&	&	\\\hline
BTmag &		Mean BT magnitude   						& + &	\\\hline
e\_BTmag &	Standard error on BTmag   					& + &	\\\hline
VTmag &		Mean VT magnitude   						& + &	\\\hline
e\_VTmag &	Standard error on VTmag   					& + &	\\\hline
m\_BTmag &	Reference flag for BT and VTmag   			& 	&	\\\hline
B\_V &		Johnson B-V colour   						& + & +	\\\hline
e\_B\_V &	Standard error on B-V   					& + & +\\\hline
r\_B\_V &	Source of B-V from Ground or Tycho   		&	&	\\\hline
V\_I &		Colour index in Cousins' system   			& + & +	\\\hline
e\_V\_I &	Standard error on V-I   					& + & +	\\\hline
r\_V\_I &	Source of V-I   							&	&	\\\hline
CombMag &	Flag for combined Vmag, B-V, V-I   			&	&	\\\hline
Hpmag &		Median magnitude in Hipparcos system   		& + &	\\\hline
e\_Hpmag &	Standard error on Hpmag   					& + & +\\\hline
Hpscat &	Scatter on Hpmag   							& + &   +\\\hline
o\_Hpmag &  Number of observations for Hpmag   			& ~ & ~ \\\hline
m\_Hpmag &	Reference flag for Hpmag   					& ~ & ~ \\\hline
Hpmax &		Hpmag at maximum (5th percentile) 			& ~ & ~ \\\hline
HPmin &		Hpmag at minimum (95th percentile) 			& ~ & ~ \\\hline
Period &	Variability period (days) 					& + & +\\\hline
HvarType &	Variability type   							& ~ & ~ \\\hline
moreVar &	Additional data about variability   		& ~ & ~ \\\hline
morePhoto &	Light curve Annex   						& ~ & ~ \\\hline
CCDM &		CCDM identifier   							& ~ & ~ \\\hline
n\_CCDM &	Historical status flag   					& ~	& ~ \\\hline
Nsys &		Number of entries with same CCDM   			& ~ & ~ \\\hline
Ncomp &		Number of components in this entry   		& ~	& ~ \\\hline
MultFlag &	Double/Multiple Systems flag   				& ~ & ~ \\\hline
Source &	Astrometric source flag   					& ~ & ~ \\\hline
Qual &		Solution quality   							& ~ & ~ \\\hline
m\_HIP &	Component identifiers   					& ~ & ~ \\\hline
theta &		Position angle between components   		& ~ & ~ \\\hline
rho &		Angular separation between components   	& ~ & ~ \\\hline
e\_rho &	Standard error on rho   					& ~ & ~ \\\hline
dHp &		Magnitude difference of components   		& ~ & ~ \\\hline
e\_dHp &	Standard error on dHp   					& ~ & ~ \\\hline
Survey &	Flag indicating a Survey Star   			& ~ & ~ \\\hline
Chart &		Identification Chart   						& ~ & ~ \\\hline
Notes &		Existence of notes   						& ~ & ~ \\\hline
HD &		HD number {\textless}III/135{\textgreater}  & ~ & ~ \\\hline
BD & Bonner DM {\textless}I/119{\textgreater}, {\textless}I/122{\textgreater}   & ~ & ~ \\\hline
CoD &		Cordoba Durchmusterung  (DM) {\textless}I/114{\textgreater} & ~ & ~ \\\hline
CPD &		Cape Photographic DM {\textless}I/108{\textgreater}   & ~ & ~ \\\hline
V\_I\_red &	V-I used for reductions   					& ~ & ~ \\\hline
SpType &	Spectral type \footnote{the list of used fields in decoded form is given in the Appendix A}  											& + & + \\\hline
r\_SpType &	Source of spectral type   						& ~ & ~ \\\hline
\enddata
\end{deluxetable*}
\newpage

\section{} \label{sec:appC}

\startlongtable

\begin{deluxetable*}{llcc}
	\tabletypesize{\footnotesize}
	\tablewidth{0pt}
	\tablecaption{Used parameters from the paper of Floor van Leeuwen\label{tab:fvl}}
	\tablehead{
		\colhead{Name} &
		\colhead{Description from the } &
		\colhead{Used in the full set} &
		\colhead{Used in a set}\\
		\colhead{} &
		\colhead{paper of Floor van Leeuwen} &
		\colhead{} &
		\colhead{with a restriction of}\\
		\colhead{} &
		\colhead{} &
		\colhead{} &
		\colhead{observational selection}
	}
	\decimalcolnumbers
	\startdata
	HIP &	Hipparcos identifier 					&		&	\\\hline
	Sn &	[0,159] Solution type new reduction 	&		&	\\\hline
	So &	[0,5] Solution type old reduction 		&		&	\\\hline
	Nc &	Number of components 					&	+ 	& ~ \\\hline
	RArad &		Right Ascension in ICRS, Ep=1991.25 &	~	&	~	\\\hline
	DErad &		Declination in ICRS, Ep=1991.25 	&	~	&	~	\\\hline
	Plx &		Parallax 							&	+ 	&	~	\\\hline
	pmRA &	Proper motion in Right Ascension 		&	+ 	&	~	\\\hline
	pmDE &	Proper motion in Declination 			&	+ 	&	~	\\\hline
	e\_RArad &	Formal error on RArad 				&	+ 	&	+\\\hline
	e\_DErad &		Formal error on DErad 			&	+ 	&	+\\\hline
	e\_Plx &		Formal error on Plx 			&	+ 	&	+\\\hline
	e\_pmRA &		Formal error on pmRA 			&	+ 	&	+\\\hline
	e\_pmDE &		Formal error on pmDE 			&	+ 	&	+\\\hline
	Ntr &		Number of field transits used &	+ &	~	\\\hline
	F2 &		Goodness of fit 					&	+ &	+\\\hline
	F1 &		Percentage rejected data &	+ &	+\\\hline
	var &		Cosmic dispersion added &	~	&	~	\\\hline
	ic &		Entry in one of the suppl.catalogues &	~	&	~	\\\hline
	Hpmag &		Hipparcos magnitude &	+ &	~	\\\hline
	e\_Hpmag &		Error on mean Hpmag &	+ &	+\\\hline
	sHp &		Scatter of Hpmag &		+ &	+\\\hline
	VA &		[0,2] Reference to variability annex &	~	&	~	\\\hline
	B-V &		Colour index &	+ &	~	\\\hline
	e\_B-V &	Formal error on colour index 		&	+ &	+	\\\hline
	V-I &	V-I colour index 						&	+ &		\\\hline
	UW &	Upper-triangular weight matrix\footnote{each element of the matrix is represented as a separate parameter} & + & + \\\hline
	\enddata
\end{deluxetable*}

\bibliography{biss_hip}{}
\bibliographystyle{aasjournal}
\end{document}